\documentclass[aps,twocolumn,showpacs,floats,superscriptaddress]{revtex4-1}
\usepackage{graphicx}
\usepackage{dcolumn}
\usepackage{bm}
\usepackage{color}
\usepackage{amssymb}

\begin{document}
\author{B. Capogrosso-Sansone}
\affiliation{Institute for Theoretical Atomic, Molecular and Optical Physics,
Harvard-Smithsonian Center of Astrophysics, Cambridge, MA, 02138}

\title{Solid phases and pairing in a mixture of polar molecules and atoms}

\begin{abstract}
We consider a mixture of hard core bosonic polar molecules, interacting via
repulsive dipole-dipole interaction, 
and one atomic bosonic species. The mixture is confined on a two-dimensional
square lattice and, at low enough temperatures, can be described by the two-component Bose-Hubbard
model.  The latter displays a extremely rich phase diagram including solid, superfluid, supersolid phases. 
Here we mainly focus on the checkerboard molecular solid, 
stabilized by the long range dipolar interaction, and study how the presence of atoms affects its robustness
both at zero and finite temperature. We find that, due to
atom-molecule interaction, solid 
phases can be stabilized at both, (much) lower strengths of dipolar interaction
and higher temperatures, than when no atoms are present. As a byproduct, atoms also order in a solid phase
with
same melting temperatures as for molecules. Finally, we find that for large
enough
interaction between atoms and molecules a paired supersolid phase can be
stabilized.
\end{abstract}

\maketitle
Recent years have witnessed spectacular advances in atomic and molecular
physics. Ultracold bosonic atoms have been successfully loaded in one-~\cite{Esslinger}, two-~\cite{Porto}
three-dimensional~\cite{Greiner} optical lattices with full control over the hamiltonian
parameters. The Mott-insulator (MI) to superfluid (SF) transition has been
experimentally observed and extensively studied theoretically~\cite{Bloch,Lewenstein}. Moreover, more
recently,
new experimental efforts towards trapping two component atomic mixtures in
optical lattices have been put forward~\cite{Minardi1,Minardi2,Schneble,Sengstock,Weld}. In the presence of a second component, 
new fascinating phenomena and exotic quantum phases which cannot be
accessed with single species gases, become available~\cite{Svistunov-Duan,Soyler}.
\\ \indent 
Ultracold polar
molecules, interacting via long range dipolar interaction, also represent a
fertile ground for the study of novel exotic phases of matter. The dipole-dipole
interaction can be tuned and shaped via external static and microwave electric
fields~\cite{Micheli}. Self-assembled dipolar crystal, dipolar Bose-Einstein condensate,
supersolids, are among the phases which can be stabilized with cold polar
molecules~\cite{Buchler-Baranov,BCS}. Besides, interesting applications in fundamental physics, e.g.
measurement of permanent electron dipole moment, in quantum chemistry and
few-body physics, and in quantum information processing have been envisioned~\cite{Carr}.
\\ \indent 
Observation of many body quantum phases requires high phase space
density of molecules in stable ground states. Cooling high density samples of molecules to quantum
degeneracy has proved to be challenging due to their complex internal state structure. 
Only very recently samples of \textit{high density} polar and non-polar molecules in the rovibronic
ground-state and a selected hyperfine state
were realized~\cite{Jila1,Nagerl}. The Jila group has also reported on observations of
dipolar collisions and evidence of spatial anisotrpy of dipolar interaction in a 
sample of KRb molecules with non zero induced dipole moment, i.e. aligned by external electric field~\cite{Jila2}. The
Innsbruck group has produced a high density sample of $\rm{Cs_2}$
molecules in their rovibronic ground state
trapped into an optical lattice~\cite{Nagerl}. The procedure used in~\cite{Nagerl} 
can be generalized to heteronuclear molecules.
\\ \indent In the present work we consider a novel system constituing of a
mixture of polar molecules and one
atomic, non reactive species. This system is realizable experimentally since residual atoms naturally occur 
after molecule formation is completed and can be selectively removed~\cite{Jila3}.
We are interested in studying how the presence of the atomic species affects
molecular solid
phases stabilized by dipolar interaction~\cite{BCS}. Hence we consider the case where
only reactive atoms are removed from the system, and both, molecules
and atoms, are in the absolute (including hyperfine) ground state, so that
inleastic collisions are forbidden. Examples of non reactive mixtures include
(KRb+Rb), (LiCs+Cs), (RbCs+Cs)~\cite{Cote-Hutson}.
\\ \indent 
We consider bosonic
polar molecules in a static electric field along the \textit{z}-direction which
induces a dipole moment $d$ parallel to the field. Molecules interact via
dipole-dipole interaction $V^{\rm{dd}}(R)=\frac{d^2}{4\pi\epsilon_0}\frac{R^2-3z^3}{R^5}$.
If the system is confined on the \textit{xy} plane (by e.g. a strong optical lattice potential
along the \textit{z}-drection) in order to avoid instabilities which might arise
due to the attractive part of the dipolar interaction, the resulting effective
2D potential is isotropic, $V^{\rm{dd}}_{\rm eff}\sim \frac{d^2}{r^3}$. An additional 
2D optical lattice with spacing \textit{a} (which we use as unit length) further confines the system.
If atoms are present, they interact with molecules
via Van der Waals (anisotropic) interaction, i.e. $V_{\rm{VdW}}(r)=\frac{C_6}{r^6}$, which we describe 
via an \textit{effective} $s$-wave scattering length, and via dipole-dipole
interaction $V^{\rm{dd}}_{\rm{at-mol}}(r)=\frac{C_3}{r^3}$. Since atoms have small polarizabilities,
dipolar interaction between atoms and molecules can be neglected, provided the electric field is small.
Indeed, for DC electric fields $E_{DC}\sim $ kV/cm, and  molecular (induced) dipole moment $d\sim1D$, the dipolar interaction at
a typical lattice spacing distance $a\sim 500 $~nm is $V^{dd}_{\rm{at-mol}}(a)\sim 0.1-1$~Hz $h$ 
(to be compared with molecule-molecule dipolar interaction $V^{dd}(a)\sim$~kHz $h$).
At low enough temperature, the system is described by the two-component
Bose-Hubbard Hamiltonian: 
\begin{eqnarray}
& & H = -J_{b} \sum_{<i,j>} b^{\dag}_{i} b^{}_{j} -J_{a} \sum_{<i,j>}
a^{\dag}_{i} a^{}_{j}
+ V\sum_{i<j}\frac{n_{i}^{b} n_{j}^{b}}{r^3_{ij}}\nonumber \\  
& & -U_{ab} {\sum}_{i}  n^{a}_i  n^{b}_i  
+ \frac{1}{2}{\sum}_{i} U_{aa}\; n^{a}_i
( n^{a}_i-1) \nonumber \\
& & - \sum_i \mu^{b} n^{b}_i - \sum_i \mu^{a} n^{a}_i\;\; ,
\label{hamiltonian}
\end{eqnarray}
where $b_{i}$ ($a_{i}$) and $b^{\dag}_{i}$ ($a^{\dag}_{i}$) are bosonic
operators
for molecules (atoms) with $b^{\dag 2}_{i}=0$ (the hardcore limit is well
justified if one starts with low enough molecular density so that there are no doubly occupied sites~\cite{Buchler}), and $n^{a,b}_{i}$
is the occupation number operator; the first and second terms in
Eq.~\ref{hamiltonian} describe the standard kinetic energy with hopping rate
$J_{a,b}$ for atoms and molecules respectively, the third term describes the
repulsive dipole-dipole interaction between molecules with $r_{ij}=|i-j|$,
the fourth and fifth terms describe the atom-molecule 
and atom-atom contact interaction, respectively; $\mu_{a,b}$ is the
chemical potential.   
\\ \indent 
We study model~(\ref{hamiltonian}) via large scale Monte-Carlo simulations by the Worm Algorithm~\cite{Worm,Soyler}. 
An Ewald summation is used in the interaction potential so that no cutoff in the range of the dipole-dipole interaction is introduced. 
The complete phase diagram is extremely rich. It includes
mixtures of a molecular solid or supersolid (SS) with MI or SF of the atomic
component, double SF phases, etc. Besides, one expects the presence of paired phases which
have been proved to exist in multicomponent atomic systems (see e.g.~\cite{Svistunov-Duan,Soyler}). 
In the present work we mainly focus on the checkerboard (CB) molecular solid phase at density $n_b= 0.5$~\cite{BCS}. 
Our results can be summarized as follows. Due to atom-molecule interaction, the CB molecular solid
appears at weaker dipolar interaction compared to the purely molecular
sample. Correspondingly, solid melting temperatures are higher than
in the absence of atoms. As a byproduct, atoms are also ordered in a CB solid phase
with same melting temperatures as for molecules. We also find that, for large
enough interaction between atoms and molecules a \textit{paired supersolid} (PSS)
phase, i.e. a SS phase of a composite (atom+molecule) object, can be stabilized upon doping of the solid.
\begin{figure}
\begin{center}
\includegraphics[width=\columnwidth]{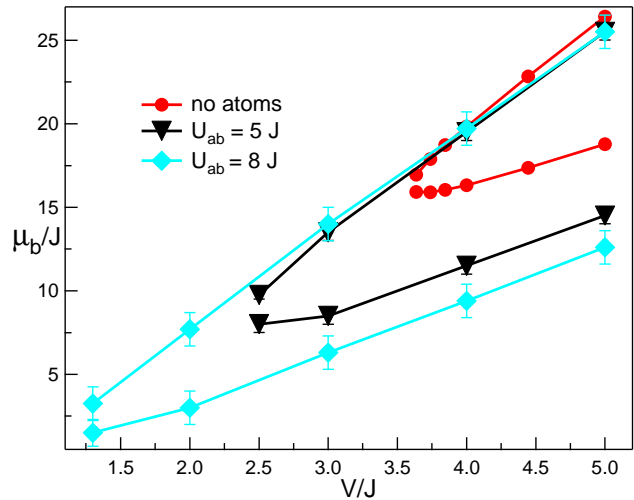}%
\end{center}
\caption{(Color online) Checkerboard lobe with parameters $n_{a}=0.5$,
$J_a=J_b$, and $U_{ab}/J =5$, 8, down traingles and diamonds respectively. 
The lobe in absence of atoms is also plotted for
reference (circles), with data taken from~\cite{BCS}. 
Errorbars are within symbol size when not visible.}
\label{fig1}%
\end{figure}
\begin{figure}
\begin{center}
\includegraphics[width=\columnwidth]{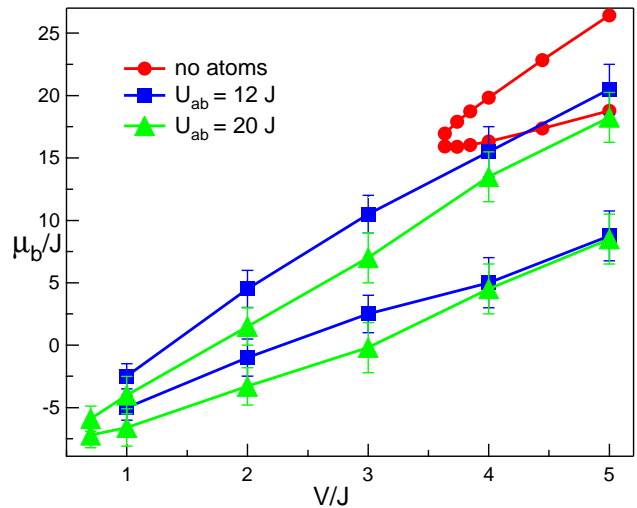}%
\end{center}
\caption{(Color online) Checkerboard lobe with parameters $n_{a}=0.5$,
$J_a=J_b$, and $U_{ab}/J =12$, 20, up traingles and squares respectively. 
The lobe in absence of atoms is also plotted for
reference (circles), with data taken from~\cite{BCS}. 
Errorbars are within symbol size when not visible.}
\label{fig2}%
\end{figure}
\\ \indent Let us first discuss the parameter space we focus on. In the following 
we consider $J_a=J_b=J$ (although the results discussed below hold in general for $J_a\sim J_b$).
This can be easily achieved by making use of two independent lattices for atoms and molecules. We work at densities 
$n_a\sim n_b\sim 0.5$, in the limit of hard-core molecules and atoms, and consider attractive 
interaction between atoms and molecules, i.e. $U_{ab}>0$. Results still hold for repulsive atom-molecule interaction, i.e. $U_{ab}<0$,
provided one replaces atoms with holes (and viceversa) in the following discussion. 
The hard-core limit is naturally reached for molecules~\cite{Buchler}, as we have mentioned above. For atoms, it
can easily be reached via Feshbach resonances. The qualitative results found here, though, are still valid 
for soft core atoms as long as $U_{aa}>|U_{ab}|\gg J$. For example, at $U_{aa}\sim2.5U_{ab}$, and $U_{ab}=8J$ no 
qualitative difference from the hard-core case was observed. This can be understood from the fact that, in the limit
$U_{aa}\gtrsim |U_{ab}|\gg J$, the difference between hard- and soft-core appears in fourth order perturbation theory. 
The constraint on particle numbers can be satisfied in the presence of an external harmonic confinement, 
which provides a scan over chemical potential. 
We have indeed observed that, when molecules form a solid, the atomic density follows the molecular one
for a wide 
range of atomic chemical potentials  and already at relatively low atom-molecule
attraction, implying that $n_a\sim n_b\sim 0.5$ will be easily satisfied in some regions of the trap.
\\ \indent
Let us now turn to the results. In Fig.~\ref{fig1} and~\ref{fig2} we study how the onset of the CB solid 
is affected by the presence of atoms, upon varying the interaction between atoms
and molecules
limiting ourselves to values $U_{ab}/J$ for which atomic density also
orders, forced by the molecular solid. 
In Fig.~\ref{fig1} we plot the CB lobes for $U_{ab}/J=5$ (down
triangles), $U_{ab}/J=8$
(diamonds), and, for reference, in absence of atoms (circles). 
(We do not investigate the nature of phases outside the lobe, as establishing the full phase 
diagram of model~(\ref{hamiltonian}) is beyond the scope of this work). 
The boundaries of the lobes are determined by monitoring $n_b(\mu_b)$ for system sizes L=12, 16, 20, 24, 30 
(with extrapolation to infinite size when necessary).
With no atoms in the system the solid
is stabilized for dipolar interaction strength $V/J\gtrsim 3.6$. In the presence of atoms, instead, the solid already exists for 
$V/J\gtrsim 2.5\; (1.75)$ at atom-molecule interaction $U_{ab}/J=5\; (8)$. 
Similarly, Fig.~\ref{fig2} shows  CB lobes for $U_{ab}/J=12$ (squares), $U_{ab}/J=20$ (up triangles), and in
absence of atoms (circles). Here the effect of the atom-molecule interaction is even more dramatic. The
solid is stabilized for $V/J\gtrsim 0.5$, a factor of 7 smaller than the case of no atoms present in the system! 
The renormalization of the tip of the lobe towards smaller values of $V/J$ can be
understood by noticing that, for $U_{ab}/J$ values considered here, atoms also ordered in a
CB solid.
One can therefore visualize a solid of a composite object, constituent of one atom and one molecule,
which has an effective hopping $J_{\rm eff}\propto \frac{J^2}{U_{ab}}<J$ responsible of 
rescaling the onset of CB solid towards smaller dipolar interaction strengths.
\\ \indent The results summarized in Fig.s~\ref{fig1}~and~\ref{fig2} have important consequences
for experiments, as they show that in presence of non reactive atoms,
\textit{smaller} induced dipole moment is needed in order to stabilize the solid. 
If one considers (RbK+Rb) mixtures loaded in lattices of depth $V_{a}\sim8E_r^{a}$ and $V_{b}\sim7E_r^{b}$ ($E_r^{\alpha}$ 
is the recoil energy of species $\alpha$) and with spacing $a\sim500$nm, one has $J_a\sim J_b\sim80$Hz~$h$ and $U_{ab}/J\sim8$,
under the assumption $-a_s^{\rm{KRb-Rb}}\sim150 a_0$. 
In absence of atoms, observation of CB solid requires dipolar interaction $V\gtrsim 3.6 J\sim288$Hz~$h$ 
which can barely be reached with fully polirized molecules. In the presence of atoms, instead, 
one needs $V\gtrsim 1.5 J\sim120$Hz~$h$, which requires an induced dipole moment $d\gtrsim0.35D$ (easily achievable
in the laboratory with alkali dimers).
\\ \indent
The enhanced robustness of the CB solid further manifests itself in larger melting temperatures
compared to the case of no atoms present. In Fig.~\ref{fig3} we plot the critical temperature for the melting of the CB solid $T_c/J$
as a function of $V/J$ for $U_{ab}/J=5$ (diamonds), 8 (down triangles), 12 (squares),
20 (up triangles), and with no atoms (circles). We find that critical temperatures have already saturated at $U_{ab}/J=8$
and can be up to a factor of 2 larger than in absence of atoms. (We have not resolved $T_c/J$ close to the tip of the 
lobes where simulations are more challenging.) Within our statistical errors, melting temperatures for the atomic density wave 
are the same as for molecules, suggesting that a solid of a composite (atom+molecule) object survives up to the melting transition line.
To appreciate the effect of atoms let's consider an induced dipole moment of 0.5 D in a (RbCs+Cs) mixture loaded in a lattice 
with spacing $a=400$ nm and depths $V_{a(b)}\sim9(7)E_r^{a(b)}$. With $-a_s\sim150 a_0$ we obtain $J_a\sim J_b\sim70$Hz~$h$, $U_{ab}/J\sim12$, 
and a solid melting temperature 2 times larger than in absence of Cs atoms. 
\begin{figure}
\begin{center}
\includegraphics[width=\columnwidth]{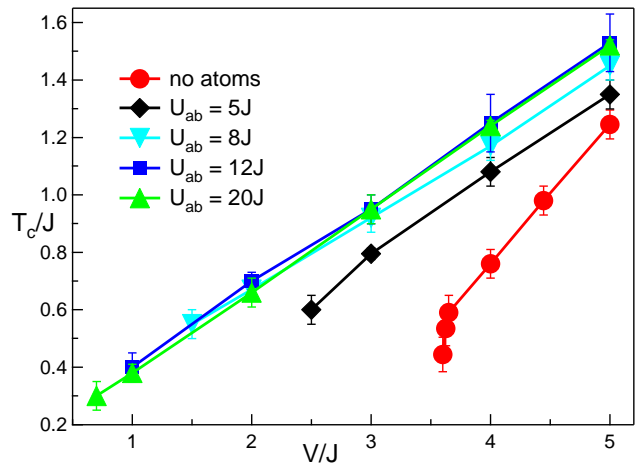}
\end{center}
\caption{(Color online) Checkerboard solid melting temperatures with parameters $n_a=n_b=0.5$, $U_{ab}/J=5$, 8, 12, 20 
diamonds, down triangles, up triangles, squares respectively and, for reference, in absence of atoms (cricles).}
\label{fig3}%
\end{figure}
\\ \indent
Next we turn to the discussion of paired phases. Here, we would like to focus on the
possibility of realizing a paired supersolid (PSS), i.e. a SS of a composite object formed by an atom and a polar molecule
(we remind the reader that a SS is a state of matter featuring both superfluid and crystalline order).
In order to check for the existence of a PSS we have calculated the structure factor and the winding number statistics, 
upon doping the CB solid with interstials, i.e. we work at densities $n_a\sim n_b<0.5$. The structure factor S(\textbf{k}), 
which characterize solid phases, is defined as:
\begin{equation}
S_{\alpha}(\textbf{k})=\sum_{\textbf{r},\textbf{r}'}
\exp\left[i\textbf{k}\left(\textbf{r}-\textbf{r}'\right)\right]\langle
n_{\alpha}(\textbf{r}) n_{\alpha}(\textbf{r}')\rangle/N_{\alpha}
\end{equation}
with $\alpha=a,b$, $\textbf{k}=(\pi,\pi)$ the reciprocal lattice
vector for the CB solid, and $N_{\alpha}$ the total number of particles of species $\alpha$. Paired superfluidity (PSF), 
i.e. superfluidity of the composite object, is instead identified by simultaneuos finite statistics of the \textit{sum} of winding numbers squared
$<(W_a+W_b)^2>\;\propto n_s^{\rm{PSF}}$, where $n_s^{\rm{PSF}}$ is the superfluid stiffness, and zero
statistics of the \textit{difference} of winding numbers squared $<(W_a-W_b)^2>$. We have found that, in order to observe PSF, one needs 
interactions $U_{ab}/J\gtrsim12$. The relatively small effective hopping $J_{\rm{eff}}\propto\frac{J^2}{U_{ab}}$, 
or equivalently large effective mass, associated with PSF, and the presence of long range interaction
renders the observation of a PSS a fairly challenging task. On one hand the superfluid stiffness is indeed suppressed 
by the enhancement of the effecticve mass of the composite object, on the other hand $V/J_{\rm{eff}}\gg1$ gives rise to metastability~\cite{BCS}. 
Since we have found that the CB solid remains robust against temperature and doping also close to the tip of the lobe, the optimal situation for 
the observation of PSS is indeed in this region of the phase diagram, where metastability is much less severe.
\begin{figure}
\begin{center}
\includegraphics[width=\columnwidth]{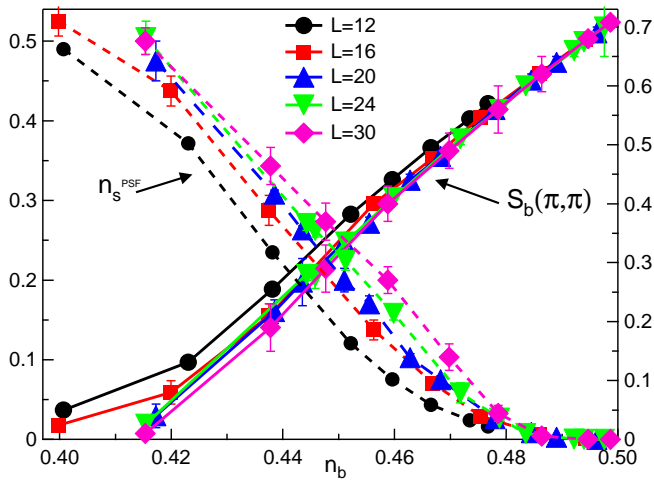}%
\end{center}
\caption{(Color online) Structure factor $S_b(\pi,\pi)$ (right axis) and pair superfluid stiffness $n_s^{\rm{PSF}}$ 
(left axis) as a function of $n_b$ with parameters $J_a=J_b$, $n_a=n_b$, $U_{ab}/J=12$, $V/J=0.5$, $\beta=L/J$, 
and linear system L=12 (circles), 16 (squares), 20 (up triangles), 24 (down triangles), 30 (diamonds). 
Both $S_b(\pi,\pi)$ and $n_s^{\rm{PSF}}$ are non zero for a finite range of densities $n_b<0.5$, signifying the existence of a PSS phase. 
Errorbars are within symbol size when not visible.}
\label{fig4}%
\end{figure}
\\ \indent 
In Fig.~\ref{fig4} we plot $S_b(\pi,\pi)$ (right axis) and   
$n_s^{\rm{PSF}}$ (left axis) as a funciton of density with parameters $J_a=J_b$, $n_a=n_b$, $U_{ab}/J=12$, $V/J=0.5$, 
linear system sizes L=12 (circles), 16 (squares), 20 (up triangles), 24 (down triangles), 
30 (diamonds), and inverse temperature $\beta=L/J$. There exist a finite range of densities for which 
both $S_b(\pi,\pi)$ and $<(W_a+W_b)^2>$ are finite. (We have checked that $S_a(\pi,\pi)$ and $<(W_a-W_b)^2>$ are simultanuesly 
finite and zero respectively.) While for $S_b(\pi,\pi)$ there are not finite size effects within errorbars and provided $L>12$, 
for $n_s^{\rm{PSF}}$ finite size effects are present. As L increases so does $n_s^{\rm{PSF}}$, signifying 
that in the thermodynamic limit $n_s^{\rm{PSF}}$ is indeed finite. Pronounced size effects, which can be seen even at densities $n_b\sim0.43$, 
where the CB solid is disappearing, are most likely due to low Kosterlitz-Thouless (KT) temperatures for the PSF-normal transition. This is further 
jutified by finite temperature results. We have indeed found that at $n_a=n_b\sim0.45$, $T_{\rm{KT}}\lesssim 0.05 J$ (not shown here).
\\ \indent 
The large effective mass of the composite object enhance metastability. As a result it is computationally challenging
to establish how far the PSS phase extends. We have found the presence of many metastable phases already at $V/J\sim1$
and $n_b\lesssim0.5$, not very far from the tip. These states are characterized by the presence of various grain boundaries and/or extended defects. 
These observations are also relevant in the context of PSS
with polar molecules in bilayered systems with no inter-layer tunneling~\cite{Trefzger}. 
In~\cite{Trefzger} the authors study soft core bosons with in-plane \textit{nearest neighbor} repulsion and inter-layer
\textit{nearest neighbor} attraction. By using a combination of mean field Gutzwiller approach and exact diagonalization to study an 
effective hamiltonian of pairs, they predict the existence of a PSS phase. 
Polar molecules on bilayers with zero inter-layer tunneling can be interpreted as a two-component mixture with same intra-species long range interactions
while the inter-species one, also long range, can be independently tuned (by changing the distance between layers). 
The model studied here differs from the one describing polar molecules in bilayers in that both, atom-atom and atom-molecule interactions are short range.  
We expect the metastability to become even more severe when all interactions are long range, 
as it happens in the case of bilayers, making the existence of a PSS phase questionable and surely extremely challenging to observe both in simulations 
and experimentally. Preliminary simulations validate this expectation.
\\ \indent
In conclusion, we have studied a mixture of hard core bosonic polar molecules 
and one atomic non reactive bosonic species, confined on a two-dimensional
square lattice. We have shown that the presence of atoms renders the CB molecular solid
phase more robust both at zero and finite temperature, i.e. CB forms at lower strengths of dipolar interaction
and survives at higher temperatures. We have estimated that this effect could be sizeable in current experimental setups. 
As a byproduct, atoms also order in a solid phase with
same melting temperatures as for molecules. We have also established that for large
enough
interaction between atoms and molecules a PSS can be stabilized. We expect the observation of the latter to be 
experimentally challenging due to the presence of metastability and the low temperatures at which the superfluid component disappears.
\\ \\
We would like to especially thank E. Kuznetsova for enlightining discussions. We also acknowledge fruitful discussions with N. Prokof'ev, G. Pupillo,
S. Rittenhouse, H. Sadeghpour, S. Soyler, T. Tscherbul. This work was supported by the Insitute
for Atomic, Molecular and Optical Physics (ITAMP).
%

\end{document}